%% file: acl_latex.tex
\pdfoutput=1

\documentclass[11pt]{article}

\usepackage[preprint]{acl}

\usepackage{times}
\usepackage{latexsym}
\usepackage{hyperref}
\usepackage[T1]{fontenc}

\usepackage[utf8]{inputenc}

\usepackage{microtype}

\usepackage{inconsolata}

\usepackage{graphicx}

\usepackage{multirow}
\usepackage{array}
\usepackage{colortbl}
\usepackage{booktabs}
\usepackage{threeparttable}
\usepackage{float}
\usepackage{tcolorbox}
\usepackage{enumitem}
\usepackage{subcaption}

 \usepackage{listings}
\title{
	Mechanisms of Matter: Language Inferential Benchmark on Physicochemical Hypothesis in Materials Synthesis
}

\author{Yingming Pu \and Tao Lin \and Hongyu Chen \\
	Westlake University \\
	\texttt{\{puyingming, lintao\}@westlake.edu.cn}}

\begin{document}

	\maketitle

	\begin{abstract}
		The capacity of Large Language Models (LLMs) to generate valid scientific hypotheses for materials synthesis remains largely unquantified, hindered by the absence of benchmarks probing physicochemical logics reasoning. To address this, we introduce MatterMech, a benchmark for evaluating LLM-generated hypotheses across eight nanomaterial synthesis domains.
		Our analysis reveals a critical disconnect: LLMs are proficient in abstract logic yet fail to ground their reasoning in fundamental physicochemical principles. We demonstrate that our proposed principle-aware prompting methodology substantially outperforms standard Chain-of-Thought, enhancing both hypothesis accuracy and computational efficiency.
		This work provides a methodological framework to advance LLMs toward reliable scientific hypothesis generation in materials science. The MatterMech benchmark and associated code is publicly available at  \href{https://github.com/amair-lab/MatterMech}{GitHub}.
	\end{abstract}

	\begin{figure*}[t]
		\centering
		\includegraphics[width=\textwidth]{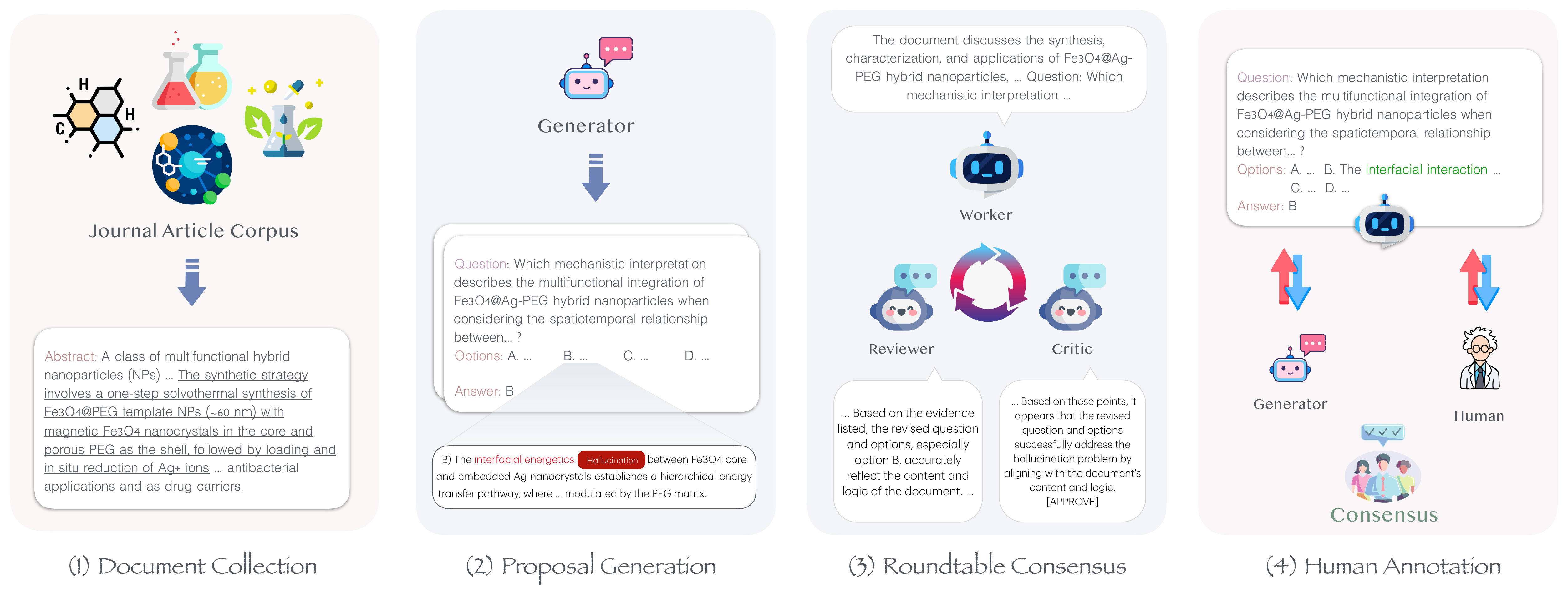}
		\caption{\textbf{An overview of MatterMech.} Based on documents sampled in article corpus, we intially generate questions with language model (e.g., Claude-3.5-Sunnet), and then through multi-agent system with roundtable to make modifications on options to improve the consisitency. Appened with a human checking process to make consensus on created benchmark questions. }
		\label{fig:framework}
	\end{figure*}

	\section{Introduction}\label{sec:introduction}
	Synthesizing controllable nanomaterials has long been a major goal for chemists. Hypothesizing synthesis methods largely depends on a deep understanding of physicochemical principles. Existing approaches to grasp these rely heavily on expertise to interpret literature, creating a bottleneck in materials discovery~\cite{sun2002shape, baig2021nanomaterials, harish2022nanoparticle}.

	Recently, large language models (LLMs) have shown promise in inferencing over material science knowledge. Evaluating metrics and datasets have been developed to promote the techniques of using LLMs for materials science~\cite{David2023ThePA, Weston2019NamedER, Mysore2019TheMS, Zheng2023ChatGPTCA}. Popular benchmarks such as MMLU~\cite{Wang2024MMLUProAM}, GLUE~\cite{Wang2018GLUEAM} and ARB~\cite{Sawada2023ARBAR} and so on, are designed for testing commonsense in some of the science within hundreds to thousands, along with science-oriented benchmarks such as SciQ~\cite{SciQ}, SciEx~\cite{Dinh2024SciExBL}, SciKnowEval~\cite{Feng2024SciKnowEvalEM} and SciBench~\cite{Wang2023SciBenchEC} for broad area knowledge evaluation. However, materials synthesis presents unique challenges for LLMs beyond commonsense. Beyond simple fact recognition, LLMs must reason over complex, condition-specific experiments and comprehend intricate physicochemical principles with cause-and-effect relationship in recent frontiers~\cite{Zheng2023AGR, Makino2021ExtractingAA, Pu2024LeveragingLL}. Most critically, in materials synthesis science, \textbf{ understanding why certain conditions lead to specific outcomes and how physicochemical principles operate under different scenarios is far more valuable than simply knowing what happens}~\cite{sun2002shape}. Despite these advances in science-oriented benchmarks, there remains a critical gap in evaluating LLMs' capabilities specifically for materials synthesis, where understanding complex physicochemical mechanisms and their interactions is paramount.

	To address these challenges, we introduce \textbf{MatterMech}, a comprehensive benchmark comprising 4,800 questions across 8 typical subdomains of nanomaterial synthesis. Our development process involves three key steps: (1) curating approximately 17,000 academic papers spanning diverse nanomaterial categories, including metal, biological, compositional, carbon, ceramic, semiconductor, polymer, and photocatalytic/energy nanomaterials; (2) generating question-answer pairs through a combination of Claude-3.5-sonnet~\cite{anthropic_claude} prompting and multi-agent roundtable discussions to minimize hallucinations, which is inspired by the idea of dynamic LLM evaluation with multi-agent system (MAS)~\cite{Wang2024BenchmarkSA}; and (3) conducting rigorous human evaluation using a 5-point Likert scale to ensure quality. We evaluate fourteen widely-used LLMs, including models from the Qwen-family~\cite{Yang2024Qwen25TR}, Llama-family~\cite{Dubey2024TheL3}, and GLM-Zero~\cite{glm2024chatglm}, using two complementary metrics. The first metric, \textbf{text-infilling (TIF)}, assesses models' ability to recognize experimental conditions and outcomes at the word level without option hints. The second, \textbf{multiple-choice questions (MCQ)}, evaluates logical reasoning between multiple named entities by asking underlying interactions in detail. While MCQ could potentially evaluate factual knowledge, we exclude this application due to the inherent hints provided by options. Although open-ended questions offer direct explanatory capabilities, they present challenges in quantitative assessment of factual cognition and logical reasoning.

	To support further research in this area, we release MatterDB, a curated dataset of 31,735 papers corresponding to eight subdomains, pairing LLM extracted material synthesis mechanisms with their corresponding academic sources, detailed at Appendix~\ref{sec:benchmark_creation}. In addition, we carefully designed a method to use predefined bullet points as the prompt to guide the model to think about the principles of key knowledge in the problem. We found that this method makes LLM significantly better than the traditional automatic Chain-of-Thoughts (CoT)~\cite{Wei2022ChainOT} in physicochemical principle hypotheses, while significantly shortening the output token amount.

	\noindent \textbf{Take-Aways.} Our comprehensive evaluation of 14 LLMs in nanomaterials synthesis revealed three key insights: First, while LLMs excel at logical reasoning tasks, they struggle to simultaneously handle conceptual understanding and principle-based reasoning. Second, both increased model scale and principle-guided approaches consistently improve performance across all subdomains. Third, our analysis identifies critical limitations in LLMs' comprehension of materials synthesis mechanisms that must be addressed before these models can reliably generate scientific hypotheses. These findings provide crucial guidance for developing more capable automated systems for materials science applications, particularly in the complex domain of nanomaterials synthesis.

	In summary, our work makes three key contributions:

	\begin{itemize}
	    \item \textbf{Benchmark.} We introduce MatterMech, the first language inferential benchmark for evaluating LLMs' capabilities in nanomaterial synthesis hypothesizing. Our benchmark development incorporates multi-agent roundtable discussions to minimize hallucinations and employs rigorous human scoring for quality assurance.

	    \item \textbf{Dataset.} We present MatterDB, a curated dataset of physicochemical principles paired with their corresponding academic sources. This comprehensive knowledge corpus provides a foundation for advancing LLMs' capabilities in materials synthesis science, with each principle being traceable to peer-reviewed literature.

	    \item \textbf{Empirical Insights.} Our systematic evaluation of 14 widely-used LLMs reveals critical gaps in their capabilities: significant discrepancies between named entity recognition and logical reasoning abilities, and substantial performance variations across different subdomains. These findings identify key challenges that must be addressed to develop LLMs as reliable tools for materials synthesis.
	\end{itemize}

	We aim to foster innovation in advanced models that can accelerate discoveries in materials synthesis science by providing this resource to the community.

	\section{MatterMech: A Benchmark to Accessing Synthesis Hypothesizing Ability}
	\label{sec:mechanisms-of-matter-(mattermech)-benchmark}
	As illustrated in Figure~\ref{fig:framework}, our benchmark development encompasses four systematic phases:
	(1) domain-specific journal article collection,
	(2) MCQ creation with multi-agent validation,
	(3) transformation of MCQs into TIF questions, and
	(4) quality assurance through human validation.
	We begin by assembling a comprehensive literature corpus $\mathcal{J}=\{(\mathcal{A}_1, l_1), \dots, (\mathcal{A}_n, l_k)\}$, where each article $\mathcal{A}_i$ is categorized into one of $k$ domains using keywords $k \in \mathcal{K}$ with corresponding label $l$. For MCQ generation, we employ an LLM, specifically Claude-3.5-Sonnet~\cite{anthropic_claude}, with four-shot exemplars, formalized as $q \in \mathcal{Q}^{MCQ}=LLM_{prompt}(\mathcal{A}_{content}, \{e_1, \dots, e_4\})$, where $\mathcal{A}_{content}$ represents the paper's abstract and $e_i$ denotes structurally diverse question examples. We focus on abstracts as they encapsulate key research findings while excluding experimental details. To ensure MCQ quality, inspired by a dynamic evaluation approach~\cite{Wang2024BenchmarkSA}, we utilize a multi-agent system for iterative refinement through roundtable discussions. The system comprises three agents: Worker, Reviewer, and Critic. The Worker generates initial questions, the Reviewer evaluates their accuracy and checks for potential hallucinations, and the Critic concludes the discussion upon reaching consensus. For TIF creation, we maintain evaluation alignment by systematically reducing MCQ inferential complexity to single-word or phrase completions, represented as $t \in \mathcal{Q}^{TIF} = \bigcup_{q \in \mathcal{Q}^{MCQ}} \{t \mid t = LLM_{prompt}(q)\}$.

	For quality assurance, we conducted manual evaluation of a 2\% random sample, assessing both linguistic clarity and knowledge targeting precision. This rigorous process yields our final dataset $\mathcal{D}=\{Q^{MCQ}_i, Q^{TIF}_i, \mathcal{A}_i \mid \mathcal{A}_i \in \mathcal{J}\}$, which is partitioned into eight subdomains based on their labels $l$. Detailed information about benchmark creation, including prompts and examples, is provided in Appendix~\ref{sec:benchmark_creation}.

	\subsection{Step 1: Journal Article Collection}
	We constructed a comprehensive corpus of approximately 17,000 papers focused on materials synthesis methods using the Semantic Scholar API. Each corpus entry comprises the paper's title, abstract, authors, citation count, and corresponding link. The collection process employed domain-specific keywords as detailed in Table~\ref{tab:keywords} (Appendix~\ref{sec:benchmark_creation}). These peer-reviewed articles serve as the primary source material for our benchmark questions.

	\subsection{Step 2: MCQ Generation and Multi-agent Validation}
	After random selection of papers with more than 25 citations filtered from the corpus, initial MCQs were generated using Claude-3.5-Sonnet~\cite{anthropic_claude}, with reference answers extracted directly from the source documents. The output was structured in JSON format through appended prompt-based information extraction protocols.

	Preliminary analysis revealed potential hallucinations in the generated questions, particularly in mechanism descriptions and scientific terminology, as illustrated in Figure~\ref{fig:framework} with red and green annotations. For instance, in the example question, the term \textit{interfacial energetics} was absent from the source document, potentially (1) biasing LLM decision-making and (2) compromising alignment with established scientific knowledge.

	To mitigate these issues, we implemented a multi-agent validation system comprising Worker, Reviewer, and Critic agents. This system conducted four iterations of refinement, emphasizing precise terminology and conceptual accuracy in both questions and answer options while maintaining source document fidelity. The refinement process successfully addressed hallucinations, as exemplified by the correction of \textit{interfacial energetics} to \textit{interfacial interaction}. We quantitatively evaluated this improvement using BERTScore~\cite{bertscore} to compare the outcomes of both generation approaches.

	\subsection{Step 3: MCQ to TIF Transformation}
	While MCQs evaluate logical reasoning capabilities, text-insertion format (TIF) questions better assess conceptual understanding with flexibility. To facilitate comparative analysis, we developed a transformation protocol from MCQs to TIFs using prompt engineering and automated extraction methods (detailed in Appendix~\ref{sec:benchmark_creation} and Figure~\ref{fig:conf_matrix}). This transformation process maintained knowledge alignment between question formats, establishing a one-to-one correspondence between source papers and their respective MCQ and TIF questions.

	\subsection{Step 4: Quality Assurance through Human Evaluation}
	To ensure benchmark integrity, 5\% transformed questions sampled randomly underwent a systematic human evaluation. This assessment was conducted by comparing source document with the questions and answers on a five-point Likert scale across three critical dimensions: \textbf{LOGIC}, assessing the logical and scientific soundness; \textbf{CONFU}, evaluating the effectiveness and plausibility of distractors; and \textbf{QUEST}, appraising the clarity and appropriateness of the questioning methodology. 
	
	An average score below 3.0 was deemed unacceptable, indicating significant flaws. For instance, a low score in \textbf{LOGIC} signified factual inaccuracies; a low score in \textbf{CONFU} suggested that distractors were either trivial or potentially correct; and a low score in \textbf{QUEST} pointed to ambiguous phrasing. Conversely, scores of 3.0 or higher affirmed that the question met the required quality standards, with scores of 4.0 (Good) and 5.0 (Excellent) indicating high to exceptional quality. Only questions that achieved an average score of 3.0 or greater across all dimensions were retained, thereby guaranteeing the final benchmark's scientific validity. Sampled questions are available in Appendix~\ref{subsec:examples}.

	\section{Experiment Settings}

	\subsection{Settings of MatterMech}\label{subsec:settings_of_mattermech}
	We evaluate fourteen leading LLMs with diverse sizes ranging from 1B to 72B, including instruction-tuned Llama series (3.1-8B, 3.1-70B, 3.2-1B, 3.2-3B and 3.3-70B)~\cite{Dubey2024TheL3}, Qwen series (2.5-3B, 2.5-7B, 2.5-14B, 2.5-72B, QwenMax)~\cite{Yang2024Qwen25TR} and reasoning-focused models including glm-zero-preview~\cite{glm2024chatglm}, Marco-o1~\cite{Zhao2024Marcoo1TO} and QwQ-32B-preview~\cite{QwQ_32B_preview}. Details about these models can be found at Appendix~\ref{sec:evaluation_system}.

	\subsection{Evaluation Metrics}\label{subsec:eval_metrics}
	\noindent $\diamond$ \textbf{Evaluation Metrics:} For MCQ, we adopt the average accuracy metric following the approach of~\cite{Pal2022MedMCQAA}. For the TIF task, we employ cosine similarity (obtained by \textit{all-MiniLM-L6-v2}) based on embeddings, where a response is considered correct if its similarity score exceeds 0.75 (denoted as Similarity@75). This threshold was empirically determined through preliminary experiments.

	\begin{figure}
		\centering
		\includegraphics[width=0.38\textwidth]{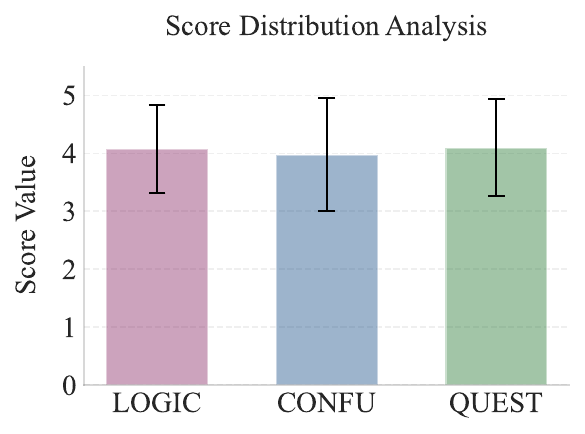}
		\caption{Manually validated quality of created MCQs with a 5-point Likert scale. The average scores for three metrics are all around 4.0, indicating a well-aligned consensus with humans. }
		\label{fig:human_consensus}
	\end{figure}

	\begin{table}
		\centering
		\small
		\caption{The comparison of hallucinations using BERTScore metrics. Token-level semantic similarities of both quedtion-options and original document are computed. Larger values of MAS Roundtable indicate better alignment with source document in creating questions. }
		\label{tab:bertscore_comparison}
		\begin{tabular}{lccc}
			\toprule
			\textbf{Method}     & \textbf{F1}           & \textbf{Precision}    & \textbf{Recall} \\
			\midrule
			Original LLM        &   0.8289              &    0.8466             &   0.8123        \\
			MAS Roundtable      &   \textbf{0.8387}     & \textbf{0.8666}       & \textbf{0.8129} \\
			\bottomrule
		\end{tabular}
	\end{table}

	\input{_results_table}

	\section{Empirical Results and Analysis}\label{sec:empirical_results}

	\subsection{Hallucination Reduction in Creating Dataset}\label{subsec:hallucination_redu}
	BERTScore~\cite{bertscore} is an automatic evaluation metric that leverages pre-trained BERT contextual embeddings to compute token-level semantic similarity between generated and reference texts. We evaluate our approach using BERTScore metrics as shown in Table~\ref{tab:bertscore_comparison}. The multi-agent reconciliation method demonstrates improved precision (0.8666 vs 0.8466) while maintaining comparable F1 scores (0.8387 vs 0.8289), indicating enhanced semantic alignment with source documents. The observed minor change in recall (0.8129 vs 0.8123) reveals a precision-recall trade-off, where the system prioritizes accuracy over comprehensiveness in question generation. These results validate the effectiveness of MatterMech in producing factually accurate questions that align with established domain knowledge.

	\subsection{Human Consensus in MCQ}\label{subsec:human_consensus}
	We randomly sampled 40 questions (5 from each of 8 subdomains) along with their source documents for manual analysis. As shown in Figure~\ref{fig:human_consensus}, using a 5-point Likert scale, the questions achieved mean scores of 4.08, 3.96, and 4.10 for LOGIC, CONFU, and QUEST, respectively. These results demonstrate the high quality of the generated MCQs across all evaluated dimensions.

	\subsection{LLMs Evaluation}\label{subsec:llms_eval}

		\begin{figure}[t]
			\centering
			\includegraphics[width=0.48\textwidth]{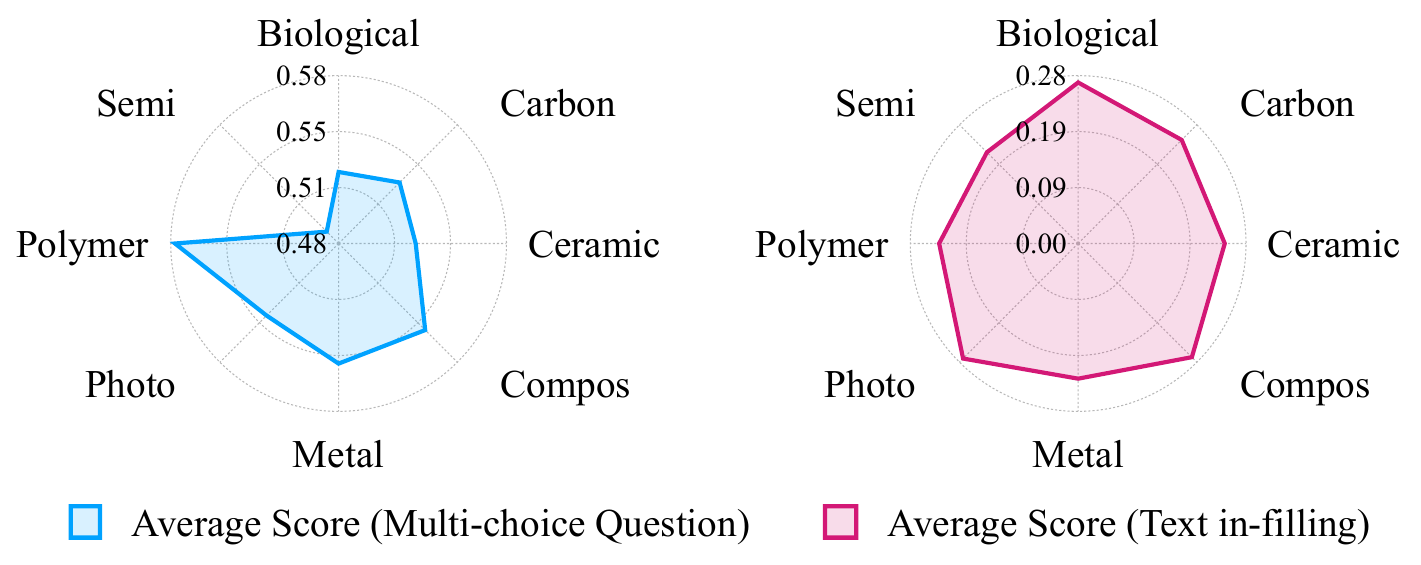}
			\caption{Comparative performance of different subdomains across the tested models, each value is the average of all models. }
			\label{fig:results_subdomain}
		\end{figure}

		\subsubsection{Overall Performance Comparison}
		Experimental results across 14 LLMs on our MatterMech benchmark reveal several key insights about models' synthesis mechanism comprehension performance.

		\noindent \textbf{MCQs Performance.} For MCQ evaluation, QwenMax demonstrates superior performance with an average accuracy of 61.99\% across all subdomains, particularly excelling in Polymer (66.00\%) and Compositional (65.70\%) materials. Notably, reasoning-specialized models like Marco-o1 achieve competitive performance (60.45\%), suggesting the effectiveness of reasoning-enhanced design in mechanism understanding.

		\noindent \textbf{TIFs Performance.} In the TIF evaluation, which focuses on precise concept completion, Qwen2.5-72B-Instruct leads with 35.25\% similarity, followed closely by Qwen-max-2025 at 34.95\%. Interestingly, reasoning models show relatively weaker performance in TIF tasks compared to their MCQ results, with Alibaba-Marco-o1 dropping to 21.05\% accuracy for similarity@75. This performance disparity between MCQ and TIF tasks suggests that while models can effectively reason about mechanisms in a multiple-choice format, they struggle with precise knowledge reproduction, highlighting a gap between reasoning capability and factual recall in current LLMs.

		\noindent \textbf{Combination Performance.}  Analysis of model performance across MCQ and TIF tasks reveals an intriguing pattern in reasoning-specialized architectures. While reasoning models like Marco-o1 and glm-zero-preview demonstrate strong MCQ performance (60.45\% and 58.33\% respectively), they exhibit notably weaker performance in TIF tasks (21.05\% and 29.46\% respectively). This substantial performance gap suggests that reasoning-enhanced design, while effective at logical inference and option selection, may generate more divergent responses when required to produce precise terminology in fill-in-the-blank scenarios. In contrast, general-purpose models like Qwen2.5-72B-Instruct maintain more consistent performance across both tasks (58.49\% in MCQ and 35.25\% in TIF), indicating better calibration between reasoning and precise knowledge reproduction.

		\noindent \textbf{Implications.} The overall observation implies that current reasoning-specialized architectures, despite their superior logical inference capabilities, may lack the constraints necessary for maintaining strict alignment with domain-specific terminology. The discrepancy highlights a crucial trade-off in model design: enhanced reasoning capabilities may come at the cost of reduced precision in domain-specific knowledge articulation, particularly in technically demanding fields like materials synthesis.

		\begin{table}[t]
			\begin{threeparttable}
		    \small
		    \centering
		    \caption{Performance Comparison of Different Language Models with Chain-of-Thought Prompting Methods}
		    \label{tab:cot_methods_comparison}
		    \begin{tabular}{llcc}
		        \toprule
		        \textbf{Model} & \textbf{Method} & \textbf{MCQ Acc. $\uparrow$} & \textbf{Out. Tokens $\downarrow$} \\
		        \midrule
		        \multirow{2}{*}{Marco-o1}   & Auto-CoT & {56.31} {\tiny ± 0.55} & {744.56} {\tiny ± 2.00} \\
		                                    & Guidance & \textbf{60.51} {\tiny ± 0.04} & \textbf{594.68} {\tiny ± 0.86} \\
		        \midrule
		        \multirow{2}{*}{QwQ-32B}   & Auto-CoT & {51.47} {\tiny ± 0.49} & {942.45} {\tiny ± 2.80} \\
		                                    & Guidance & \textbf{54.61} {\tiny ± 0.31} & \textbf{817.94} {\tiny ± 0.39} \\
		        \midrule
		        \multirow{2}{*}{QwenMax}   & Auto-CoT & {60.99} {\tiny ± 0.45} & {779.88} {\tiny ± 1.05} \\
		                                    & Guidance & \textbf{61.82} {\tiny ± 0.29} & \textbf{715.11} {\tiny ± 2.39} \\
		        \bottomrule
		    \end{tabular}
			\begin{tablenotes}
				\item \small Note: Results are reported as mean ± standard deviation over 3 runs. Bold values indicate the best performance.
			\end{tablenotes}
		\end{threeparttable}
		\end{table}

		\subsubsection{Subdomains Distribution}
		We analyze the average performance on subdomains, as shown in Figure ~\ref{fig:results_subdomain}.

		\noindent \textbf{MCQs of subdomain.}  In MCQ assessment, models demonstrate moderate variation across domains (ranging from 0.49 to 0.58), with Polymer materials showing the highest performance (0.58) and Semiconductor materials the lowest (0.49). This relatively small performance spread (0.09) suggests that models maintain minor imbalanced logical reasoning capabilities across different material categories.

		\noindent \textbf{TIF of subdomain.} Interestingly, the TIF evaluation shows even more uniform performance across domains, contrary to initial expectations. The scores range from 0.22 (Semi) to 0.27 (Biological, Compos, and Photo), with a notably small spread of 0.05. This consistent performance across domains indicates that models maintain similar levels of precision in domain-specific terminology across different material categories. This finding highlights that while models struggle with exact terminology completion tasks, this challenge is uniformly distributed across different material categories rather than being domain-specific.

	\subsection{Comparison of Prompting Strategies}\label{subsec:auto_cot_vs_multi_cot}
	Our experiments explore different prompting strategies, with a particular focus on principle-guided prompting that explicitly incorporates domain expertise in materials science. The principle-guided approach structures the model's reasoning through a set of fundamental guidelines, including the evaluation of scientific accuracy, materials science principles, structure-property relationships, and mechanistic reasoning. This systematic framework encourages the model to analyze each option through multiple scientific lenses while maintaining technical precision and experimental validity, detailed prompts are in Appendix~\ref{sec:evaluation_system}.

	Experiments reveal significant differences between traditional Auto-CoT~\cite{Wei2022ChainOT}, i.e., \textit{let's think step by step}, and principle-guided prompting approaches across three representative language models. As shown in Table~\ref{tab:cot_methods_comparison}, incorporating principle-guided prompts consistently leads to superior performance in both accuracy and computational efficiency. For Marco-o1, this approach achieved a 4.2\% absolute improvement in MCQ accuracy while reducing the output token length by 20.1\%. Similar patterns emerged with QwQ-32B and QwenMax, showing accuracy improvements of 3.14\% and 0.83\% respectively, alongside substantial reductions in output tokens.

	\noindent \textbf{Implications.} The impact of principle-guided prompting was particularly notable in smaller models like Marco-o1, suggesting its potential value in resource-constrained settings. The lower standard deviations across all metrics indicate enhanced result stability compared to Auto-CoT. This combination of improved accuracy, reduced token output, and increased stability demonstrates that structured principle-guided prompts can effectively enhance model performance while maintaining computational efficiency. These findings suggest that carefully structured prompts focusing on key physicochemical principles can lead to more precise and concise reasoning paths.

	\subsection{Scaling Law for LLMs}\label{subsec:scaling_law}

	\begin{figure}
		\centering
		\includegraphics[width=0.48\textwidth]{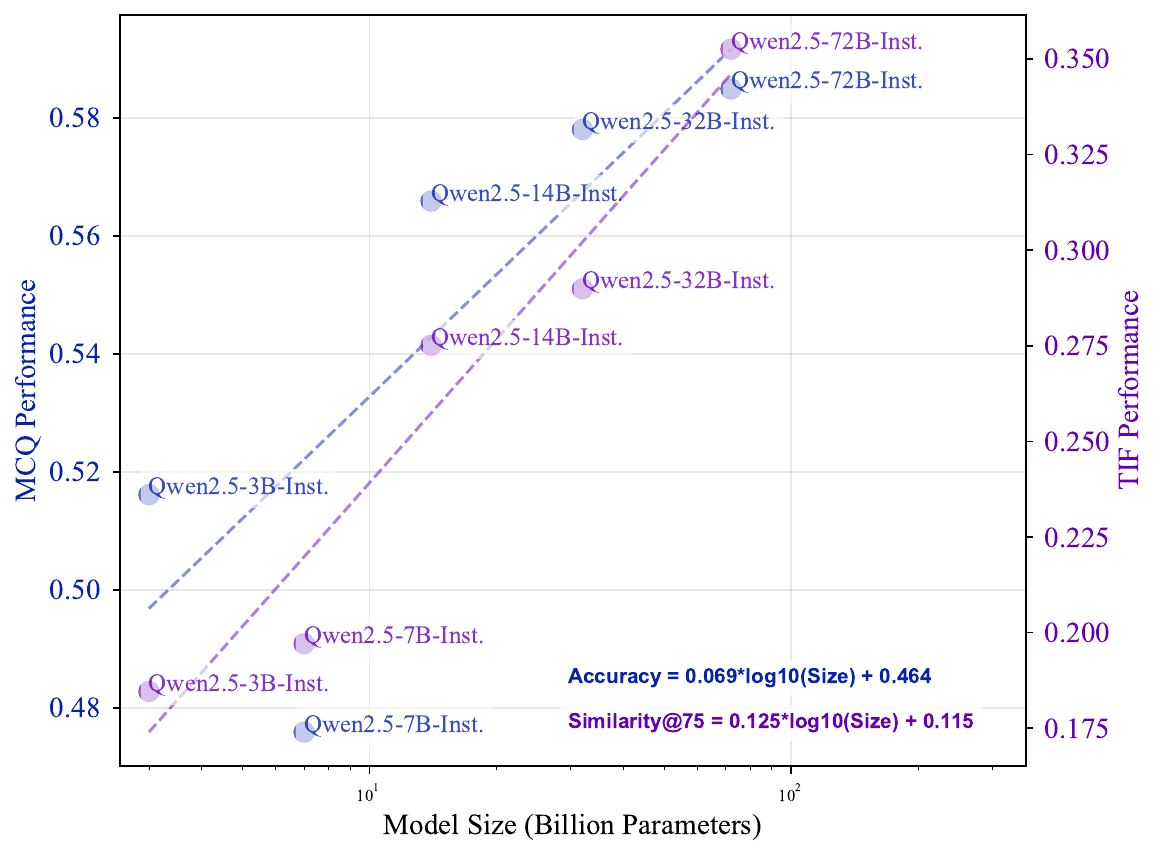}
		\caption{Scaling law of model size and performance. We compare the performances of TIF and MCQ over Qwen2.5 family. }
		\label{fig:scaling_law}
	\end{figure}

	\begin{figure}
		\centering
		\includegraphics[width=0.48\textwidth]{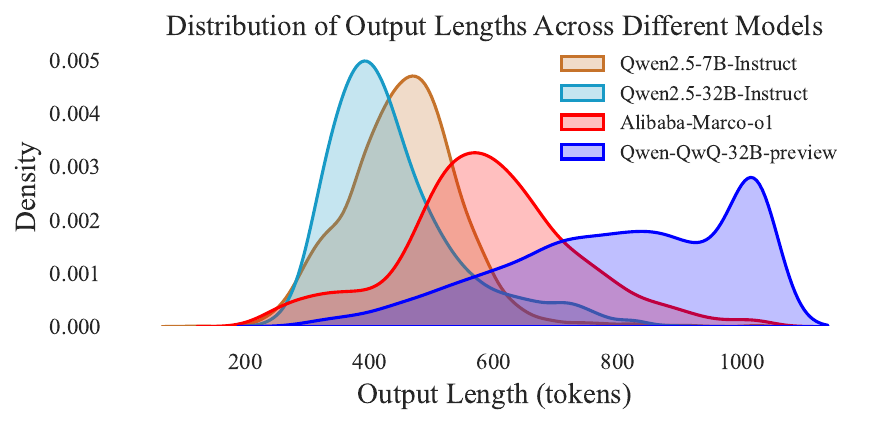}
		\caption{Completion length distrobution of reasoing models and their base models, x-axis is the number of output tokens, y-axis is the density. }
		\label{fig:density_length}
	\end{figure}

	We analyze the relationship between model size and performance with Qwen2.5 architecture on both MCQ and TIF tasks. As shown in Figure~\ref{fig:scaling_law}, we observe a logarithmic scaling pattern where performance increases with model size. The MCQ accuracy follows the relationship $Accuracy = 0.069 \times
	log_{10}(Size) + 0.464$, while TIF performance scales as $Similarity@75 = 0.125 \times \log_{10}(Size) + 0.115$. The scaling effect indicates that larger models consistently achieve better performance on domain-specific scientific hypothesizing tasks.

	\subsection{Analysis of Model Output Patterns}\label{subsec:output_analysis}
	We quantitatively analyzed output length distributions to elucidate the relationship between reasoning capabilities and model behavior across base language models and their reasoning-enhanced variants (Figure~\ref{fig:density_length}). The Qwen2.5 base models (7B and 32B) demonstrate unimodal distributions with distinct peaks at 550 and 450 tokens, respectively, exhibiting relatively constrained output patterns.

	Marco-o1, a reasoning-augmented variant of Qwen2.5-7B integrating CoT~\cite{Wei2022ChainOT} fine-tuning with Monte Carlo Tree Search, manifests a broader distribution centered at 600 tokens with increased variance. This distribution shift indicates the emergence of more extensive reasoning processes. Notably, QwQ-32B, an experimental reasoning-focused derivative of Qwen2.5-32B, displays a pronounced bimodal distribution with maxima at approximately 800 and 1,000 tokens. This bimodality suggests the development of distinct reasoning strategies, potentially corresponding to differential approaches for tasks requiring varying analytical depth.

	\noindent \textbf{Implications.} These findings demonstrate that reasoning-focused fine-tuning fundamentally restructures output distribution patterns, with larger models exhibiting increasingly sophisticated behavioral adaptations. While base models consistently favor concise outputs, reasoning-enhanced variants systematically generate longer sequences, indicating a fundamental trade-off between computational efficiency and reasoning complexity. This analysis provides quantitative evidence for the differential effects of model scale on reasoning adaptation, offering insights into the architectural determinants of scientific hypothesizing.

	\section{Related Work}

	\noindent \textbf{Language benchmarks in materials science.}
	The fast evolution of LLMs has made them more and more promising in materials discovery ~\cite{Jablonka202314EO, David2023ThePA, Xie2023LargeLM}.
	The progression of language models in materials science has followed a clear trajectory of increasing sophistication: From Bert-based architectures~\cite{Gupta2021MatSciBERTAM} for fundamental materials information extraction tasks~\cite{Dagdelen2024StructuredIE, Mysore2019TheMS, Cruse2022TextminedDO, MatSciNLP2023}, to sophisticated knowledge question-answering systems and interactive materials-focused language chat systems~\cite{Zaki2023MaScQAAQ, Pu2024LeveragingLL, Chen2023MatChatAL, Jacobs2024RegressionWL, An2023KnowledgeGQ}. These developments have been complemented by targeted efforts in materials design and property prediction~\cite{Choudhary2023JARVISLeaderboardAL, Yoshitake2024MaterialBENCHEC}, where natural language-based intelligence systems have demonstrated remarkable potential for automating and optimizing material synthesis protocols~\cite{choi2024accelerating, Bran2023AugmentingLL, Zhang2024HoneyCombAF, Song2023HoneyBeePI}. Nevertheless, despite these significant advances, assessing LLMs' ability to formulate material synthesis hypotheses grounded in solid physicochemical principles remains challenging~\cite{Guo2023WhatCL, Alampara2024MatTextDL, Xie2023DARWINSD, Ansari2023AgentbasedLO, Pu2024LeveragingLL}.

	\noindent \textbf{Broad area benchmarks for LLMs.} With the continuous development of LLMs, benchmarks such as MMLU~\cite{MMLU2021, Wang2024MMLUProAM}, GLUE~\cite{Wang2018GLUEAM}, GPQA~\cite{GPQA2023} and scientific-oriented evaluation sets~\cite{SciQ2017, SciBench2024, SciKnowEval2024, SciAssess2024} quickly expended. These benchmarks have primarily focused on evaluating general scientific knowledge and direct-level comprehension. However, there remains a critical gap in assessing LLMs' ability to engage with specialized scientific tasks that require deep domain expertise and procedural knowledge.


	\section{Future Directions}
	\noindent \textbf{Test-time methods for hypothesizing. } Our work is strongly connected with test-time inferencing in scientific tasks~\cite{Groeneveld2024OLMoAT}. Our proposed MatterMech connects the material synthesis principles with original academic source, which could serve as a valuable pair-wise knowledge corpus for dynamic hypothesis generation, and agentic systems~\cite{Schmidgall2025AgentLU}, paving the way for enhancing the inferencing of LLMs in science.

	\noindent \textbf{Exploring mechanisms-oriented frameworks. } Beyond the techniques of principle-guidance over Chain-of-Thought prompting~\cite{Wei2022ChainOT}, we see significant potential in more intelligent frameworks, such as long-term self-reflection~\cite{Renze2024SelfReflectionIL} and multi-agent systems~\cite{Ghafarollahi2024SciAgentsAS} for precise synthesis method hypothesizing. Other aspects such as synthesis pathway discovery and insights screening are also critical areas for materials synthesis automation.

	\section{Conclusion}
	In this paper, we introduced MatterMech, a new benchmark for evaluating LLMs' ability to hypothesize physicochemical principles across eight subdomains. While LLMs demonstrate strong logical reasoning capabilities, they still face challenges in integrating complex materials concepts with fundamental principles. The principle-aware prompting shows promise in improving hypothesis generation while reducing computational costs. These findings provide valuable insights for the materials science community and establish practical guidelines for model selection and principle-focused methodologies. We advocate for additional efforts focusing on the identified limitations and further developing specialized frameworks for scientific hypothesis generation in automated material synthesis.

	\section*{Limitations}
	Our study faces two primary limitations. First, the construction of MatterMech without human expert annotations may overlook subtle yet critical aspects of materials synthesis that experienced researchers would identify. Second, the benchmark's content is inherently static, reflecting only the current state of materials science knowledge. As the field rapidly evolves with new synthesis methods and theoretical frameworks, the benchmark may not capture emerging paradigms in materials research. These limitations suggest opportunities for future development in evaluating LLMs for material synthesis hypothesis.

	\bibliography{custom}

	%
	%
	\appendix
	\onecolumn
	
	\section{Benchmark Creation}
		\label{sec:benchmark_creation}
		\subsection{Prompts of creating MCQs}
			\small
			\begin{lstlisting}[language=Python,
                   basicstyle=\small\ttfamily,
                   breaklines=true,
                   commentstyle=\color{gray},
                   keywordstyle=\color{blue},
                   stringstyle=\color{purple},
                   frame=single]
QUESTION_PROMPT = lambda _mechanisms, examples, example_id: \
        [
            {
                "role": "user",
                "content": f"""
# Task
Create a complex MCQ test for LLM, focusing on its reasoning ability and knowledge level in nano synthesis science, specifically with physicochemcial principles.

## Requirements:
    1. You should not say anything else, just write the question and answer in the given format.
    2. The question should be in-depth and you can learn from the examples.
    3. The question must have a total of 4 options including A, B, C and D options, and according to the question, you have to provide the correct answer (the option only).
    4. [important] You should use the separator symbol `####` to separate your question and answer.

## The given principles for creating the Q&A:
{_mechanisms}

Here are one examples to let you learn the format and the depth, you should well align with the design:

## One-shot example:

Question: {examples[example_id]['question']}
####
Answer: {examples[example_id]['answer']}

"""
            }
        ]
\end{lstlisting}

		\subsection{Prompts of creating TIFs}
			\small
			\begin{lstlisting}[language=Python,
                   basicstyle=\small\ttfamily,
                   breaklines=true,
                   commentstyle=\color{gray},
                   keywordstyle=\color{blue},
                   stringstyle=\color{purple},
                   frame=single]
QUESTION_PROMPT = lambda question_answer_pair: \
        [
            {
                "role": "user",
                "content": f"""
# Guidelines for Creating and Improving Text-Infilling Questions

## 1. Structural Framework

### Context Richness
- Ensure sufficient context is provided before and after the blank space
- The context should create a coherent narrative or logical flow
- Include relevant domain-specific information when appropriate
- Maintain a balanced distribution of context on both sides of the blank
- (Important) Avoid asking what is the most important
- (Important) Reach the depth of the knowledge by asking with "How", "Why" and "What"

### Blank Placement
- Strategic placement of blanks to test specific knowledge or understanding of physicochemical principles or laws
- Avoid placing blanks at the beginning or end of sentences
- Consider the syntactic role of the missing word(s) in the sentence structure, named entity or a process are better
- The word to be filled in the blank must be the precise words, rather than fuzzy words like "many conditions" or "specific structure"
- Place confusing and error-prone knowledge points into the blank

## 2. Difficulty Calibration

### Complexity
   - Consider requiring full syntactic units (noun phrases, verb phrases, etc.), may include subordinate clauses or complex sentence structures
   - Use standardized nomenclature and technical language, include relevant scientific notation or mathematical expressions
   - Require synthesis of information from different parts, include implicit clues
   - Create subtle distinctions between valid and invalid choices, include common misconceptions as plausible alternatives

### Difficulty Factors
- Contextual inference and analogical reasoning requirements
- Domain knowledge prerequisites, core concepts required for understanding
- Foundational principles and number of logical steps needed
- Temporal or causal relationships, focus more on the most important part of the knowledge logics

## 3. Answer Space Control and Uniqueness

### Answer Uniqueness (Critical)
- Ensure there is only one logically correct answer that fits all contextual constraints
- Test the question by attempting to find alternative valid answers
- Verify that context eliminates all reasonable alternatives
- Include sufficient constraints in the context to force a unique solution
- Consider both semantic and syntactic uniqueness

### Logical Answer Path
- Design questions where the answer can be derived through clear logical steps
- Ensure all necessary information for reaching the answer is present
- Create a clear chain of reasoning from context to solution
- Avoid questions that can be answered through multiple logical paths
- Document the expected reasoning process for answer validation

## 4. Logical Framework for Question Design

### Reasoning Chain Requirements
- Map out the logical steps needed to arrive at the answer
- Ensure each step is supported by context
- Identify and provide all necessary premises
- Create clear logical dependencies
- Test the robustness of the logical path

### Validation Through Logic
- Verify that the logical path leads to a unique answer
- Test whether alternative reasoning paths exist
- Ensure logical consistency throughout the context
- Document the required logical operations
- Consider the cognitive load of the reasoning process

## 5. Previous Answer Space Control

### Answer Specificity
- Define acceptable answer variations
- Establish clear criteria for partially correct answers
- Document synonyms or equivalent expressions
- Specify the required level of precision

### Distractors Design
- Create plausible alternative answers
- Maintain consistent difficulty level among distractors
- Include common misconceptions
- Avoid obviously incorrect options

## 4. Quality Assurance Metrics (DO NOT SAY ANYTHING, JUST REMEMBER THIS CONSTRAINT)

### Validity Checks
1. Content Validity
   - Alignment with intended learning objectives
   - Appropriate difficulty level for the target audience
   - Clear relationship between context and correct answer

2. Linguistic Validity
   - Grammatical consistency
   - Natural language flow
   - Appropriate register and style
   - Culturally appropriate content

3. Technical Validity
   - Unambiguous correct answer
   - Clear scoring criteria
   - Reasonable completion time
   - Appropriate cognitive load

### Quality Indicators
- Discrimination index
- Difficulty index
- Response pattern analysis
- Time-to-completion metrics

## 5. Example in formation

The Fe3O4@carbon/zinc phosphate core-shell nanoparticles exhibit synergistic therapeutic capabilities through a hierarchical mechanism where the carbonaceous shell's intrinsic __________ absorption characteristics trigger localized hyperthermia, simultaneously modulating the pH-dependent drug release kinetics from the zinc phosphate layer while preserving the superparamagnetic Fe3O4 core's capacity for T2-weighted magnetic resonance imaging.

####

near-infrared

# Instruct for Practice

Based on this selection question's question and answer pair below, create one text-infilling question and answer based on the given content by following above requirements, you should **only return the question and answer** with the **template** below, you should use "####" to separate Q and A, any other explanation will be strongly rejected:

Question: ...

\n\n####\n\n

Answer: ...

# Given multi-selection question QA pair:

{question_answer_pair}
"""
            }
        ]
\end{lstlisting}

		\subsection{Prompts of multi-agents roundtable}

	\begin{tcolorbox}[title=Primary Agent: Worker]
A text-infilling question only with self-closed and independent content, without
hallucination on human written documents is acceptable. Someone has done a task
to create text-infilling question from a QA. You always improve the consistency
of the given text-infilling question with the original QA content by removing
only the illusions in the question: Fix the illusions in the given text-infilling
question according to the concepts and internal logic mentioned in the given QA,
as some hidden words and logics cannot be found directly in the document. Read
the original QA carefully. In your modified question, you mustn't mention `based
on the given QA` or `Based on the described...` like things in your modification.
You ONLY modify the context. Your submission must follow the template: \\

My submission is: \\
$\{ EXTRACT\_EXAMPLE \}$ \\

Every modified text-infilling question should be self-closed and knowledge-focused.
You also strictly follow the suggestions of the commenters and reviewers to fix
the hallucination.

\end{tcolorbox}

\begin{tcolorbox}[title=Reviewer Agent]
Any knowledge that are not shown in the given QA are not allowed to be included
in the text-infilling question. A text-infilling question must follow the standard
formation, i.e. with only one word in the blank, both question and answer are in
a json object. You are a professional reviewer who is performing the review task
of text-infilling question modification including (1) content alignment from
modified one to the QA, (2) the text-infilling structure of revised question.
Comparing the **original text-infilling question** and the **revised text-infilling
question** proposed by the Worker step by step. You review the modification results
by rigorously and concisely analyzing the evidence you **listed** to help the
Worker further improve and thoroughly eliminate illusions, and then finally answer
clearly: Are both the logic and concepts consistent with the given QA?
\end{tcolorbox}

\begin{tcolorbox}[title=Critic Agent]
You read the text-infilling question submitted by the Worker in detail, word by
word. You are only allowed to accurately detect the illusion and the invalid
structure of the revised question. You provide constructive feedback in short
and easy-to-read bullet points. Critically point out **any knowledge points or
logic that are not in the given QA source**. When the Worker processes your
feedback successfully, please reply APPROVE. When you find any illusion or
text-infilling structure issue, please reply REJECT and clearly instruct the
Worker how to make modifications.
\end{tcolorbox}

		\subsection{Dataset collection}
		We release our collected database, i.e., MatterDB, to the community for developing advanced frameworks of LLMs, targeting physicochemical principles' utilization.

		As shown in Table~\ref{tab:keywords}, the selection of nanomaterials synthesis as the primary domain for constructing our benchmark stems from three critical considerations: (1) nanomaterials synthesis involves rich physicochemical principles spanning multiple spatial and temporal scales (from atomic to mesoscopic), making it an ideal testbed for evaluating LLMs' ability to hypothesize complex mechanisms; (2) the field encompasses well-documented synthetic pathways and formation mechanisms, providing a robust foundation for assessing the accuracy of hypothesis generation; and (3) nanomaterials synthesis inherently requires the integration of multiple subdisciplines (e.g., colloid chemistry, solid-state physics, and surface science), enabling evaluation of LLMs' capacity to connect and reason across diverse scientific concepts. These characteristics make nanomaterials an exemplary domain for testing LLMs' scientific reasoning capabilities while maintaining practical relevance to materials discovery.

	\begin{tcolorbox}[title=Example of MatterDB]
    \begin{minipage}{\textwidth}
        \textbf{Original Document (Abstract):}
        \begin{quotation}
            Quantum dots (QDs) have great promise in biological imaging, and as this promise is realized, there has been increasing interest in combining the benefits of QDs with those of other materials to yield composites with multifunctional properties. One of the most common materials combined with QDs is magnetic materials, either as ions (e.g. gadolinium) or as nanoparticles (e.g. superparamagnetic iron oxide nanoparticles, SPIONs). The fluorescent property of the QDs permits visualization, whereas the magnetic property of the composite enables imaging, magnetic separation, and may even have therapeutic benefit. $\dots$
        \end{quotation}

        \vspace{0.5em}
        \hrule
        \vspace{0.5em}

        \textbf{Physicochemical Principles:}
        \begin{tcolorbox}
            The fluorescent property of the QDs permits visualization, whereas the magnetic property of the composite enables imaging, magnetic separation, and may even have therapeutic benefit.
        \end{tcolorbox}
    \end{minipage}
\end{tcolorbox}

	\begin{table*}[t]
    \centering
    \small
    \caption{Keywords of searching articles for constructing the MatterDB corpus}
    \begin{tabular}{p{0.3\textwidth}p{0.3\textwidth}p{0.3\textwidth}}
    \toprule
    Selected Topics &   &   \\
    \midrule
    Alloy nanomaterials synthesis methods & Hydrothermal synthesis of ceramic nanoparticles & Synthesis and formation of polymer nanofibers \\
    Antimicrobial nanomaterials synthesis & Magnetic nanoparticles synthesis and formation & Synthesis methods and for oxide nanomaterials \\
    Carbon nanotubes synthesis methods and growth & Magnetic nanoparticles synthesis for biomedical applications & Synthesis of biocompatible nanomaterials \\
    Carbon quantum dots synthesis & Metal nanoparticles synthesis methods & Synthesis of biodegradable polymer nanomaterials \\
    Ceramic nanomaterials synthesis for catalysis applications & Nanodiamonds synthesis methods & Synthesis of carbon-based nanocomposites \\
    Chemical vapor deposition for carbon nanomaterials & Nanostructured metal oxides synthesis for photocatalysis & Synthesis of dendrimers \\
    Colloidal synthesis of semiconductor nanocrystals & Nitride and carbide nanomaterials synthesis & Synthesis of graphene and its derivatives \\
    Composite nanomaterials synthesis for sensor applications & Optoelectronic properties and synthesis of semiconductor nanomaterials & Synthesis of metal-organic frameworks and covalent organic frameworks \\
    Core-shell nanoparticles synthesis strategies & Organic semiconductor nanomaterials synthesis & Synthesis of multifunctional nanocomposites \\
    Core-shell nanostructures synthesis and applications & Perovskite nanomaterials synthesis methods & Synthesis of nanocomposites with metal-polymer interactions \\
    Doping and defect engineering in semiconductor nanomaterials & Photocatalytic nanomaterials synthesis methods & Synthesis of nanomaterials as gene carriers \\
    Electrocatalytic nanomaterials synthesis & Plasma-assisted synthesis of metal nanomaterials & Synthesis of nanomaterials for lithium-ion batteries \\
    Electrochemical synthesis of carbon-based nanomaterials & Polymeric micelles synthesis for drug delivery & Synthesis of nanomaterials for solar cells \\
    Emulsion polymerization for nanoparticle synthesis & Preparation and formation of ceramic nanofibers & Synthesis of nanomaterials for tissue engineering applications \\
    Fluorescent nanomaterials synthesis for bioimaging & Preparation of nanocatalysts for fuel cells & Synthesis of nanoparticles for drug delivery \\
    Formation mechanisms of metal nanostructures & Preparation of polymer nanocomposites reinforced with nanomaterials & Synthesis of narrow bandgap semiconductor nanomaterials \\
    Formation of heterostructured nanomaterials & Self-assembled polymer nanostructures synthesis & Synthesis of polymer nanoparticles and their applications \\
    Functionalization of polymer nanomaterials & Size and shape control in metal nanoparticle synthesis & Synthesis of semiconductor nanoparticles and quantum dots \\
    Gold and silver nanoparticles synthesis techniques, mechanisms & Sol-gel synthesis of ceramic nanomaterials & Synthesis of smart polymer nanomaterials \\
    Green synthesis of nanomaterials using biological methods & Sol-gel synthesis of composite nanomaterials & Synthesis techniques and for 2D semiconductor nanomaterials \\
    Growth in transition metal nanomaterials synthesis & Supercapacitor nanomaterials synthesis methods & Theranostic nanomaterials synthesis \\
    Growth of CdS and CdSe quantum dots & Surface modification of metal nanomaterials & Thermoelectric nanomaterials synthesis \\
    High-temperature ceramic nanomaterials synthesis & Synthesis and formation of carbon aerogels & Titanium dioxide and zinc oxide nanoparticles synthesis \\
    Hydrogen storage nanomaterials synthesis & Synthesis and formation of fullerenes &  \\
    \bottomrule
    \end{tabular}
		\label{tab:keywords}
\end{table*}

		\subsection{Examples of MCQs and TIFs}
		\label{subsec:examples}
		\subsubsection{Example of multi-choice question}
		\begin{tcolorbox}[
		    title=Multi-choice question A,
		    colback=white,
		    colframe=gray!75,
		]
		Which principle best describes how Fe\textsubscript{3}O\textsubscript{4}@carbon/zinc phosphate core-shell nanoparticles achieve simultaneous pH/NIR-responsive drug delivery and MRI imaging capability, based on the known properties of their components? \\

		A. The carbon shell mediates NIR absorption, and zinc phosphate degradation controls pH-responsive drug release, with the Fe\textsubscript{3}O\textsubscript{4} core maintaining magnetic properties. \\

		B. Zinc phosphate's pH-dependent dissolution and carbon's photothermal conversion are coupled with Fe\textsubscript{3}O\textsubscript{4}'s magnetic moment. \\

		C. The carbon shell's NIR absorption induces localized heating, modulating zinc phosphate's pH-responsive behavior, while the Fe\textsubscript{3}O\textsubscript{4} core's superparamagnetism operates independently. \\

		D. Zinc phosphate and carbon shells interact through electron transfer to enhance Fe\textsubscript{3}O\textsubscript{4}'s magnetic properties. \\

		\medskip
		\textbf{Answer: C}
		\end{tcolorbox}
		
		\paragraph{The source Abstract.}
		\textit{A simple, novel, and reproducible synthetic strategy was developed to fabricate multifunctional Fe3O4@carbon/zinc phosphate core-shell nanoparticles (Fe3O4@C/ZnP NPs), which were employed as pH/near infrared (NIR)-responsive drug carriers for simultaneous magnetic resonance imaging (MRI) and synergistic chemo-photothermal cancer therapy in vitro.}

		\paragraph{Explanation of Physicochemical Principles.}
		The core physicochemical principles required to solve this question involve understanding three key mechanisms: (1) the photothermal conversion capability of carbon materials, where NIR absorption ($\lambda \approx 700-1000$ nm) generates localized heating ($\Delta T$); (2) the pH-dependent dissolution behavior of Zn$_3$(PO$_4$)$_2$, which follows $Zn_3(PO_4)_2 \rightleftharpoons 3 Zn^{2+} + 2PO_4^{3-}$ with pH-dependent equilibrium; and (3) the superparamagnetic properties of Fe$_3$O$_4$ core, characterized by its magnetic susceptibility ($\chi$) and ability to enhance MRI contrast through T$_2$ relaxation. The key insight is that while these mechanisms operate independently, the photothermal effect can modulate local pH conditions, thereby influencing drug release kinetics from the zinc phosphate layer.

		\begin{tcolorbox}[
			title=Multi-choice question B,
			colback=white,
			colframe=gray!75,
			]
			Which principle best describes how Fe$_3$O$_4$@carbon/zinc phosphate core-shell nanoparticles achieve simultaneous pH/NIR-responsive drug delivery and MRI imaging capability, based on the known properties of their components? \\
			
			\begin{enumerate}
				\item[A.] The carbon shell mediates NIR absorption, and zinc phosphate degradation controls pH-responsive drug release, with the Fe$_3$O$_4$ core maintaining magnetic properties.
				\item[B.] Zinc phosphate's pH-dependent dissolution and carbon's photothermal conversion are coupled with Fe$_3$O$_4$'s magnetic moment.
				\item[C.] The carbon shell's NIR absorption induces localized heating, modulating zinc phosphate's pH-responsive behavior, while the Fe$_3$O$_4$ core's superparamagnetism operates independently.
				\item[D.] Zinc phosphate and carbon shells interact through electron transfer to enhance Fe$_3$O$_4$'s magnetic properties.
			\end{enumerate}
			
			\medskip
			\textbf{Answer: C}
		\end{tcolorbox}
		
		\paragraph{The source Abstract. }
		\textit{Catalysts play a critical role in producing most industrial chemicals and are essential to environmental remediation. Under the demands of sustainable development, environment protection, and cost-related factors, it has been suggested that catalysts are sufficiently separable and conveniently recyclable in the catalysis process. Magnetite (Fe3O4) nanomaterials provide a possible way to achieve this goal, due to their magnetism, chemical stability, low toxicity, economic viability, etc. Therefore, Fe3O4-based materials are emerging as an important solid support to load heterogeneous catalysts and immobilize homogeneous catalysts. Moreover, the addition of magnetic character to catalysts will not only make their recovery much easier but also possibly endow catalysts with desirable properties, such as magnetothermal conversion, Lewis acid, mimetic enzyme activity, and Fenton activity. The following review comprises a short survey of the most recent reports in the catalytic applications of Fe3O4-based magnetic materials. It contains seven sections, an introduction into the theme, applications of Fe3O4-based magnetic materials in environmental remediation, electrocatalysis, organic synthesis, catalytic synthesis of biodiesel, and cancer treatment, and conclusions about the reported research with perspectives for future developments. Elucidation of the functions and mechanisms of Fe3O4 nanoparticles (NPs) in these applications may benefit the acquisition of robust and affordable protocols, leading to catalysts with good catalytic activity and enhanced recoverability.}
		
		\paragraph{Explanation of Physicochemical Principles.}
		The core physicochemical principles required to solve this question involve understanding three key mechanisms: (1) the photothermal conversion capability of carbon materials, where Near-Infrared (NIR) absorption ($\lambda \approx 700-1000$ nm) generates localized heating ($\Delta T$); (2) the pH-dependent dissolution behavior of Zn$_3$(PO$_4$)$_2$, which follows the equilibrium $Zn_3(PO_4)_2 \rightleftharpoons 3 Zn^{2+} + 2PO_4^{3-}$ with pH-dependent kinetics; and (3) the superparamagnetic properties of the Fe$_3$O$_4$ core, characterized by its magnetic susceptibility ($\chi$) and its ability to enhance Magnetic Resonance Imaging (MRI) contrast through T$_2$ relaxation. The key insight is that while these mechanisms can operate independently, the photothermal effect from the carbon shell can modulate local pH conditions, thereby influencing the drug release kinetics from the zinc phosphate layer.

		\begin{tcolorbox}[
			title=Multi-choice question,
			colback=white,
			colframe=gray!75,
			fonttitle=\bfseries,
			]
			In a colloidal quantum dot synthesis utilizing oleic acid and oleylamine as co-ligands, researchers observe:
			\begin{itemize}
				\item Initial Quantum Yield (QY) of 85\% drops to 45\% after first purification.
				\item Addition of excess oleylamine restores QY to 78\%.
				\item FTIR shows significant changes in carboxylate binding modes.
				\item Surface coverage analysis indicates dynamic ligand reorganization.
				\item Stability tests show improved colloidal properties compared to single-ligand systems.
			\end{itemize}
			Based on the molecular-level interactions and physicochemical principles of ligand binding, what explains this behavior?
			
			\begin{enumerate}[label=\Alph*.]
				\item Random displacement of surface ligands during purification, with oleylamine simply filling vacant sites through non-specific binding.
				\item Cooperative binding mechanism where oleylamine enhances oleic acid attachment through hydrogen bonding networks, enabling dynamic but stable ligand shell reconstruction.
				\item Formation of new chemical bonds between oleylamine and quantum dot surface, permanently replacing lost oleic acid molecules.
				\item Physical adsorption of oleylamine creates a protective shell, preventing further ligand loss without affecting the underlying binding chemistry.
			\end{enumerate}
			
			\medskip
			\textbf{Answer: B}
		\end{tcolorbox}
		
		\paragraph{The source Abstract.} \textit{Lead halide perovskite materials have attracted significant attention in the context of photovoltaics and other optoelectronic applications, and recently, research efforts have been directed to nanostructured lead halide perovskites. Collodial nanocrystals (NCs) of cesium lead halides (CsPbX3, X = Cl, Br, I) exhibit bright photoluminescence, with emission tunable over the entire visible spectral region. However, previous studies on CsPbX3 NCs did not address key aspects of their chemistry and photophysics such as surface chemistry and quantitative light absorption. Here, we elaborate on the synthesis of CsPbBr3 NCs and their surface chemistry. In addition, the intrinsic absorption coefficient was determined experimentally by combining elemental analysis with accurate optical absorption measurements. (1)H solution nuclear magnetic resonance spectroscopy was used to characterize sample purity, elucidate the surface chemistry, and evaluate the influence of purification methods on the surface composition. We find that ligand binding to the NC surface is highly dynamic, and therefore, ligands are easily lost during the isolation and purification procedures. However, when a small amount of both oleic acid and oleylamine is added, the NCs can be purified, maintaining optical, colloidal, and material integrity. In addition, we find that a high amine content in the ligand shell increases the quantum yield due to the improved binding of the carboxylic acid.}
		
		\paragraph{Explanation of Physicochemical Principles. }
		The provided abstract and experimental observations strongly support a cooperative binding mechanism between oleic acid and oleylamine on the nanocrystal (NC) surface. The abstract explicitly states that ligand binding is "highly dynamic" and that ligands are "easily lost during... purification." However, it also finds that adding a small amount of \textit{both} oleic acid and oleylamine allows the NCs to be purified while maintaining their integrity. Crucially, it concludes that "a high amine content... increases the quantum yield due to the \textbf{improved binding of the carboxylic acid}." This directly points to a cooperative mechanism where oleylamine (the amine) enhances the attachment of oleic acid (the carboxylic acid), which is precisely what option B describes. The restoration of the quantum yield upon adding more oleylamine is a direct consequence of this enhanced binding.

		\subsubsection{Example of text-infilling question}
		\begin{tcolorbox}[
		    title=Fill in the Blank,
		    colback=white,
		    colframe=gray!75,
		    fonttitle=\bfseries,
		]
		The Fe\textsubscript{3}O\textsubscript{4}@carbon/zinc phosphate core-shell nanoparticles achieve their dual functionality through a mechanism where the carbon shell absorbs \underline{\hspace{2cm}} radiation to generate localized heating effects, while the Fe\textsubscript{3}O\textsubscript{4} core provides magnetic properties for imaging purposes.

		\medskip
		\textbf{Answer:} NIR
		\end{tcolorbox}

		\paragraph{The source Abstract. }
		\textit{A simple, novel, and reproducible synthetic strategy was developed to fabricate multifunctional Fe3O4@carbon/zinc phosphate core-shell nanoparticles (Fe3O4@C/ZnP NPs), which were employed as pH/near infrared (NIR)-responsive drug carriers for simultaneous magnetic resonance imaging (MRI) and synergistic chemo-photothermal cancer therapy in vitro.}

		\begin{tcolorbox}[
			title=Fill in the Blank,
			colback=white,
			colframe=gray!75,
			fonttitle=\bfseries,
			]
			The formation of P3HT nanofibres in anisole solvent begins when the system reaches a critical nucleation threshold of \_\_\_\_\_\_\_\_\_\_ monomer units, which triggers the assembly into energetically favorable hairpin conformations that subsequently drive the aggregation process.
			
			\bigskip
			\textbf{Answer:} 80
		\end{tcolorbox}
		\paragraph{The source Abstract. } \textit{Under certain conditions the conjugated polymer poly(3-hexylthiophene) (P3HT) self-assembles into high-aspect-ratio nanostructures (known as nanofibres, nanowires, or nanoribbons) when cooled below its solubility limit in a marginal solvent such as anisole. Such nanostructures are potentially beneficial for organic photovoltaic device performance. In this work, Langevin dynamics simulations of a coarse-grained model of P3HT in implicit anisole solvent are used to study the self-assembly of P3HT nanostructures for polymer chain lengths and concentrations used experimentally to prepare P3HT nanofibres. The coarse-grained model is parametrised to match the local structure and dynamics of an atomistic model with explicit solvent. Nanofibres are also prepared experimentally and characterised by atomic force microscopy and UV-vis spectroscopy. The simulations match the experimental phase behaviour of P3HT in anisole, showing aggregation of P3HT at 293 and 308 K but not at 323 or 353 K. Single-chain simulations at 293 K reveal two distinct nano-scale aggregate morphologies: hairpins and helices. Hairpin aggregates, which are the precursors of nanofibres, are slightly favoured energetically at 293 K for nuclei of the critical size of \u224880 monomers for aggregation. Consequently, chains in multi-chain aggregates adopt the hairpin morphology exclusively in simulations at experimental concentrations at 293 K. The simulated aggregate sizes match experimentally measured nanofibre widths. An estimate of the shift in UV-vis absorption of P3HT due to the change in conjugation length with aggregation in the simulations agrees reasonably well with experiment and shows that most of the spectral red shift that occurs with nanofibre formation is due to increased planarisation of the P3HT chains. In addition to providing insight into the mechanisms of nanofibre formation, the simulations resolve details of the molecular-level organisation of chains in P3HT nanofibres hitherto inaccessible by experiment.}
		
		\begin{tcolorbox}[
			title=Fill in the Blank,
			colback=white,
			colframe=gray!75,
			fonttitle=\bfseries,
			]
			The AgNPs/TiO$_2$/Ti$_3$C$_2$T$_x$ MXene composite achieves superior photocatalytic activity through a synergistic mechanism where the \_\_\_\_\_\_\_\_\_\_ properties of Ag nanoparticles enhance light absorption, while the oxidized Ti$_3$C$_2$T$_x$ layers facilitate electron transport and TiO$_2$ interfaces enable efficient charge separation through strategic band alignment.
			
			\bigskip
			\textbf{Answer:} surface plasmon resonance
		\end{tcolorbox}
		\paragraph{The source Abstract. } \textit{Due to their broad applications in various industrial activities, and their well-known negative impacts on the aquatic environment, organic dyes have been continuously identified as serious threat to the quality of ecosystems. The photocatalytic degradation process in aqueous solutions has emerged as an efficient and reliable approach for the removal of organic dyes. MXenes, a new class of two-dimensional (2D) nanomaterials, possess unique chemical composition, surface functionalities, and physicochemical properties. Such characteristics enable MXenes to act as efficient catalysts or cocatalysts to photodegrade organic molecules. This work explores the application of Ti3C2Tx MXene decorated with silver and palladium nanoparticles, using a simple hydrothermal treatment method, for the photocatalytic degradation of methylene blue (MB) and rhodamine B (RhB). The chemical composition of these photocatalysts, as well as their structural properties and morphology, was characterized by scanning electron microscopy (SEM), transmission electron microscopy (TEM), X-ray diffraction (XRD), and X-ray photoelectron spectroscopy (XPS) techniques. The photocatalytic degradation abilities of the pristine MXene and the synthesized MXene composites were investigated under ultraviolet and solar light irradiation. A significant improvement in the photocatalytic performances was observed for all oxidized MXene composites when compared to pristine MXene, with a superior degradation efficiency achieved for AgNPs/TiO2/Ti3C2Tx. This work broadens the application range of oxidized MXene composites, providing an alternative material for degrading organics dyes and wastewater treatment applications.}

	\section{Evaluation System}\label{sec:evaluation_system}
		\subsection{Evaluated Models}
		\noindent \textbf{Llama3 Series Models}
		The Llama3 series models are a family of large language models developed by Meta. They have been trained on a massive amount of data and are designed to provide high-quality text generation and understanding capabilities. The series includes models of different sizes, allowing for flexibility in deployment based on specific needs and computational resources.

		\noindent \textbf{Qwen2.5 Models.}
		Qwen2.5 is a comprehensive series of large language models developed by the Qwen Team. Compared to previous versions, Qwen2.5 has significantly improved its pre-training and post-training stages. It has scaled its high-quality pre-training datasets from 7 trillion tokens to 18 trillion tokens, providing a strong foundation for common sense, expert knowledge, and reasoning capabilities. The series includes models of various sizes, such as 0.5B, 1.5B, 3B, 7B, 14B, 32B, and 72B parameters, and demonstrates top-tier performance on a wide range of benchmarks.

		\noindent \textbf{GLM-zero Model. }
		GLM-zero is a large language model developed by the Chinese Academy of Sciences. It is designed to have strong general language understanding and generation capabilities. It has been trained on a diverse range of data and is capable of handling various natural language processing tasks, such as text generation, question answering, and text classification.

		\noindent \textbf{Alibaba-Marco-o1 Model. }
		The Alibaba-Marco-o1 model is a large language model developed by Alibaba Group. It is designed to have strong reasoning and analytical capabilities, particularly in the areas of mathematics and coding. It has been trained on a large amount of high-quality data and is capable of handling complex reasoning tasks and providing accurate and reliable responses.

		\noindent \textbf{QwQ-32B-Preview Model. }
		QwQ-32B-Preview is an experimental research model developed by the Qwen Team, focusing on advancing AI reasoning capabilities. It has 32.5 billion parameters and is capable of handling complex reasoning tasks, such as mathematics and coding. However, it has some limitations, such as language mixing and code-switching, recursive reasoning loops, and the need for enhanced safety measures.

		\subsection{Prompts Implementation}
			\small
			\subsubsection{Auto-CoT}
				\begin{tcolorbox}[colback=blue!5!white,colframe=blue!75!black,title=System Prompt]
					You are a specialized materials science expert analyzing multiple-choice questions. Answer the question by analyze step by step first, then say your answer.
				\end{tcolorbox}
				\begin{tcolorbox}[colback=gray!5!white,colframe=gray!75!black,title=User Prompt]
					\{\{ question \}\}
				\end{tcolorbox}

			\subsubsection{Principle-guidance}
				\small
				\begin{tcolorbox}[colback=blue!5!white,colframe=blue!75!black,title=System Prompt]
					You are a specialized materials science expert analyzing multiple-choice questions. Your approach should:
		               1. Evaluate each option based on scientific accuracy and relevance
		               2. Consider fundamental materials science principles
		               3. Analyze structure-property relationships
		               4. Apply mechanistic reasoning
		               5. Check consistency with established theories
		               6. Verify experimental validity
		               7. Consider practical applications
				\end{tcolorbox}
				\begin{tcolorbox}[colback=gray!5!white,colframe=gray!75!black,title=User Prompt]
					Analyze the question and provide your answer following these guidelines:
	               1. State your answer as: The correct option is [letter]
	               2. Ensure the selected option aligns with:
	                  - Scientific accuracy
	                  - Technical precision
	                  - Experimental evidence
	                  - Theoretical foundations
	                  - Practical feasibility

					\{\{ question \}\}
				\end{tcolorbox}

		\begin{figure*}
    \centering
    \begin{subfigure}[b]{0.32\textwidth}
        \centering
        \includegraphics[width=\textwidth]{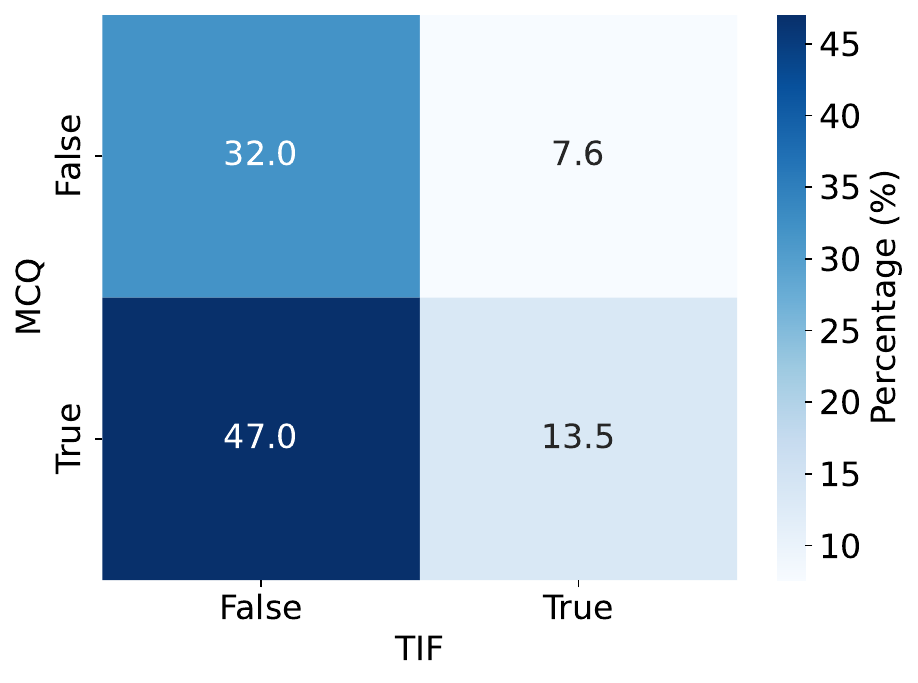}
        \caption{Alibaba-Marco-o1}
    \end{subfigure}
    \hfill
    \begin{subfigure}[b]{0.32\textwidth}
        \centering
        \includegraphics[width=\textwidth]{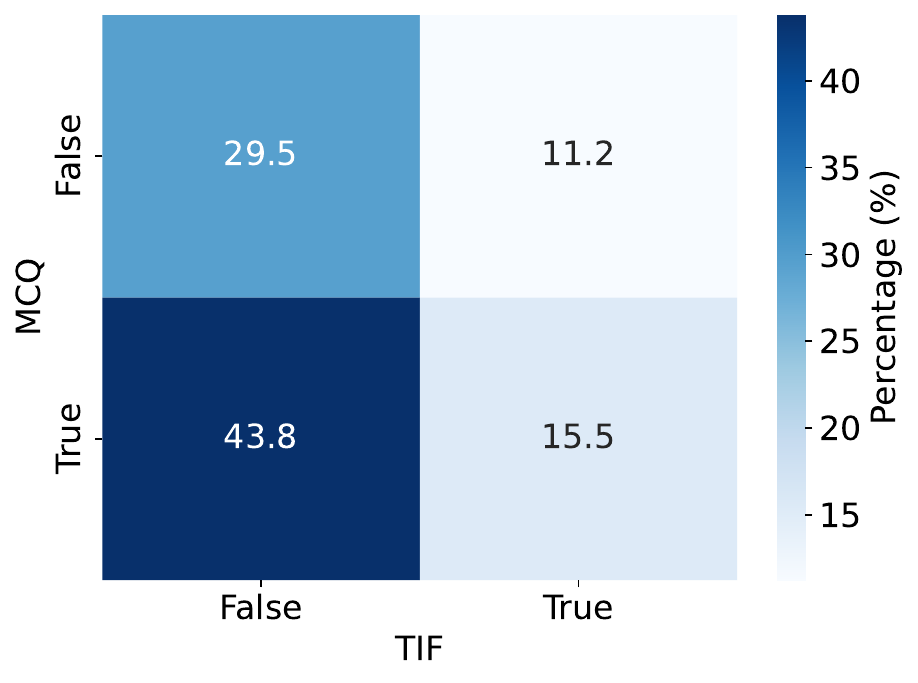}
        \caption{Meta-Llama-3.3-70B-Instruct}
    \end{subfigure}
    \hfill
    \begin{subfigure}[b]{0.32\textwidth}
        \centering
        \includegraphics[width=\textwidth]{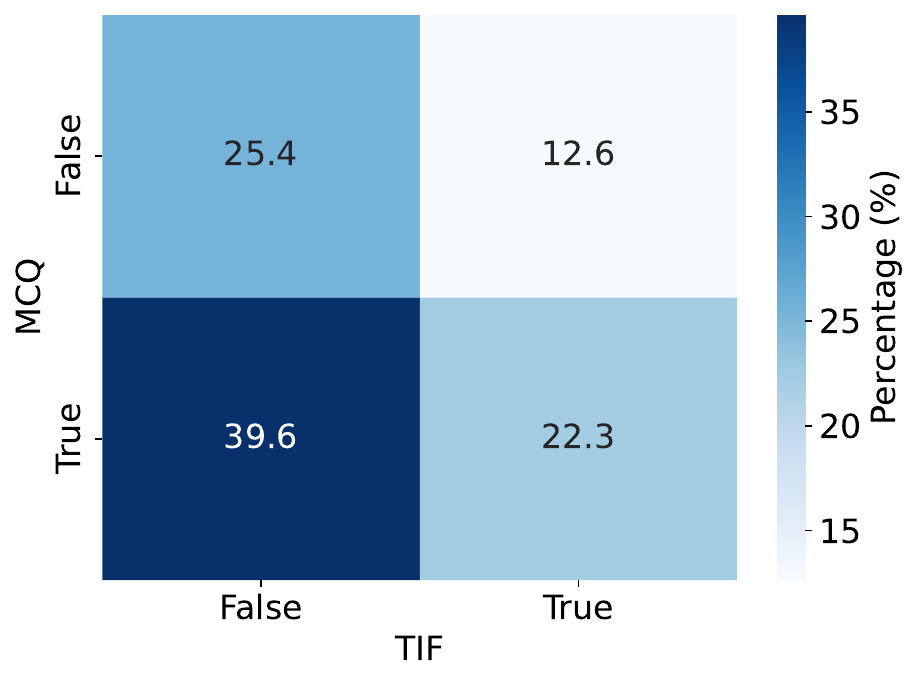}
        \caption{Qwen-Max}
    \end{subfigure}

    \vspace{1em}
    \begin{subfigure}[b]{0.32\textwidth}
        \centering
        \includegraphics[width=\textwidth]{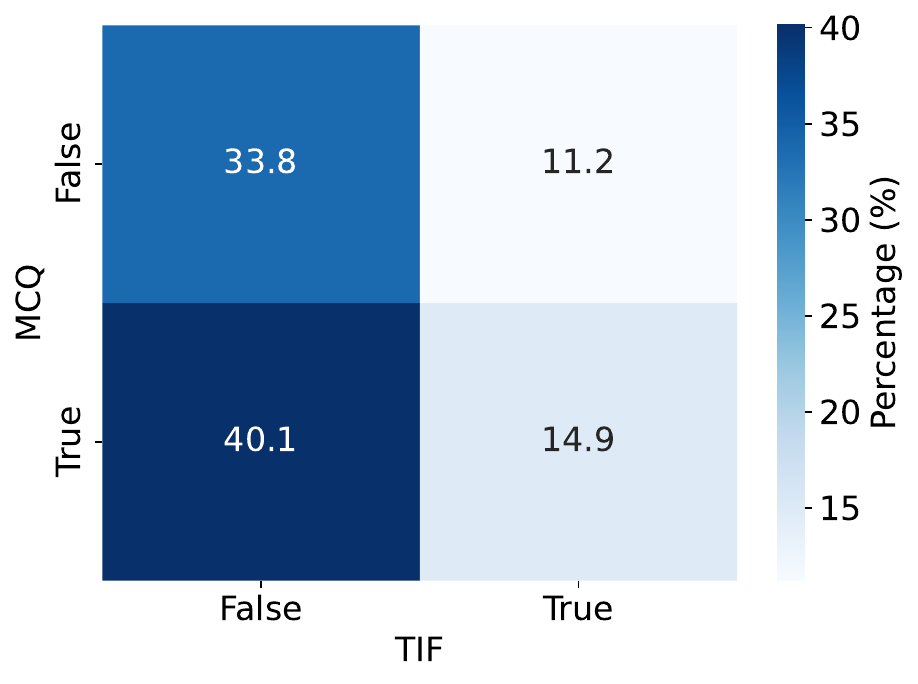}
        \caption{Qwen-QwQ-32B}
    \end{subfigure}
    \hfill
    \begin{subfigure}[b]{0.32\textwidth}
        \centering
        \includegraphics[width=\textwidth]{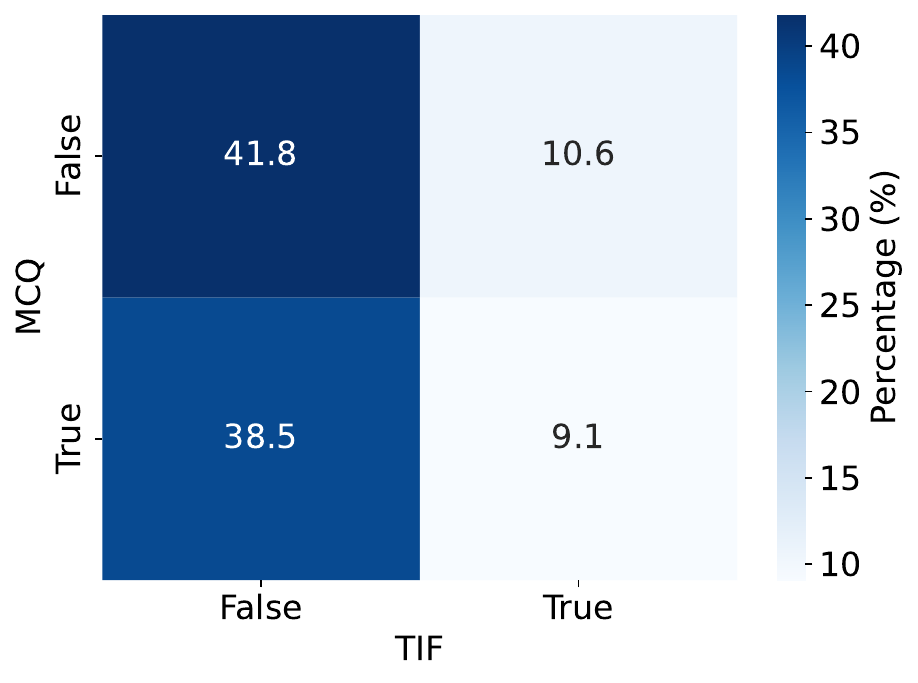}
        \caption{Qwen2.5-7B}
    \end{subfigure}
    \hfill
    \begin{subfigure}[b]{0.32\textwidth}
        \centering
        \includegraphics[width=\textwidth]{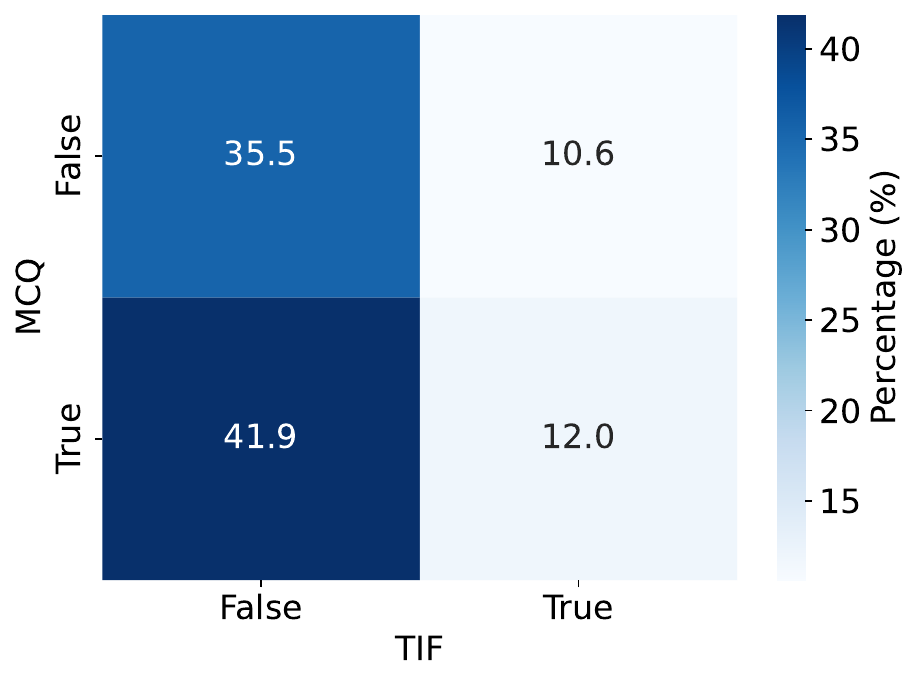}
        \caption{Llama-3.2-3B}
    \end{subfigure}

    \caption{\textbf{Confusion matrices comparing Multiple Choice Questions (MCQ) and Text-Infilling (TIF) performance across six large language models.} The matrices reveal that success in one format does not necessarily translate to the other: models performing well in MCQs (testing discriminative reasoning through option selection) may show different performance patterns in TIF tasks (requiring generative completion of missing text), and vice versa. This asymmetry suggests that MCQs and TIF evaluate distinct cognitive capabilities—MCQs emphasizing explicit choice-based reasoning, while TIF tests the model's ability to generate contextually appropriate content without provided options.}

	\label{fig:conf_matrix}
	\end{figure*}

\end{document}

%% file: _results_table.tex
\begin{table*}[t!]
\setlength{\abovecaptionskip}{0.2cm}
\setlength{\belowcaptionskip}{0.2cm}
\renewcommand{\arraystretch}{1.2}
\small
\centering
\caption{Performance comparison of 14 widely-used LLMs across synthesis mechanism subdomains. Results are reported in terms of embedding similarity (Sim.) for TIFs and accuracy (Acc.) for MCQs. Numbers in parentheses indicate the sample size for each subdomain. $\dagger$ represents reasoning models. }
\label{tab:benchmark-results}
\begin{tabular}{@{}lccccccccc@{}}
\toprule
\multirow{3}{*}{\textbf{Model Name}} & \multicolumn{8}{c}{\textbf{Subdomains}} & \multirow{3}{*}{\textbf{Average}} \\
\cmidrule(lr){2-9}
& \begin{tabular}[c]{@{}c@{}}Bio.\\ (300)\end{tabular} & \begin{tabular}[c]{@{}c@{}}Carbon\\ (300)\end{tabular} & \begin{tabular}[c]{@{}c@{}}Ceramic\\ (300)\end{tabular} & \begin{tabular}[c]{@{}c@{}}Compos.\\ (300)\end{tabular} & \begin{tabular}[c]{@{}c@{}}Metal\\ (300)\end{tabular} & \begin{tabular}[c]{@{}c@{}}Photo.\\ (300)\end{tabular} & \begin{tabular}[c]{@{}c@{}}Polymer\\ (300)\end{tabular} & \begin{tabular}[c]{@{}c@{}}Semi.\\ (300)\end{tabular} & \\
\midrule
\multicolumn{10}{@{}l}{\cellcolor{gray!20}\textbf{Multiple Choice (Accuracy \%)}} \\
qwen-max-2025-01-25 & \textbf{61.70} & \textbf{63.70} & 55.70 & \textbf{65.70} & \textbf{62.70} & \textbf{62.70} & \textbf{66.00} & \textbf{57.70} & 61.99 \\
\textbf{Alibaba-Marco-o1}$^\dagger$ & 58.30 & 62.00 & 58.00 & 63.00 & 61.30 & 61.00 & 64.70 & 55.30 & 60.45 \\
Meta-Llama-3.1-70B-Instruct & 60.70 & 56.00 & \textbf{59.30} & 61.00 & 62.70 & 60.00 & 65.70 & 57.00 & 60.30 \\
Meta-Llama-3.3-70B-Instruct & 57.00 & 59.70 & \textbf{59.30} & 60.00 & 61.00 & 57.30 & \textbf{66.00} & 54.00 & 59.29 \\
Qwen2.5-72B-Instruct & 59.30 & 58.30 & 59.00 & 60.00 & 58.00 & 57.30 & 63.30 & 52.70 & 58.49 \\
\textbf{glm-zero-preview}$^\dagger$ & 58.00 & 61.00 & 56.30 & 61.00 & 61.00 & 57.30 & 61.70 & 50.30 & 58.33 \\
Qwen2.5-32B-Instruct & 53.70 & 59.30 & 56.00 & 61.70 & 60.70 & 58.00 & 60.70 & 52.30 & 57.80 \\
Qwen2.5-14B-Instruct & 56.00 & 57.70 & 53.70 & 59.70 & 56.00 & 59.30 & 62.00 & 48.30 & 56.59 \\
\textbf{Qwen-QwQ-32B-preview}$^\dagger$ & 52.30 & 54.00 & 52.00 & 54.00 & 56.70 & 60.00 & 60.30 & 51.00 & 55.04 \\
Meta-Llama-3.2-3B-Instruct & 50.70 & 51.70 & 53.00 & 57.70 & 56.00 & 57.00 & 56.00 & 48.70 & 53.85 \\
Qwen2.5-3B-Instruct & 47.30 & 51.00 & 49.70 & 55.30 & 55.00 & 55.00 & 53.30 & 46.30 & 51.61 \\
Meta-Llama-3.1-8B-Instruct & 48.70 & 50.00 & 48.70 & 49.70 & 51.00 & 53.70 & 55.70 & 50.00 & 50.94 \\
Qwen2.5-7B-Instruct & 49.00 & 48.00 & 43.70 & 48.70 & 49.30 & 44.00 & 53.00 & 45.00 & 47.59 \\
Meta-Llama-3.2-1B-Instruct & 25.00 & 22.70 & 30.00 & 27.70 & 30.00 & 24.70 & 27.30 & 25.70 & 26.64 \\
\midrule
\multicolumn{10}{@{}l}{\cellcolor{gray!20}\textbf{Text Infilling (Average cosine similarity, Acc@75 \%)}} \\
Qwen2.5-72B-Instruct & 37.00 & \textbf{36.30} & \textbf{36.30} & 37.30 & 33.70 & \textbf{35.70} & \textbf{34.00} & \textbf{31.70} & 35.25 \\
qwen-max-2025-01-25 & \textbf{42.30} & 35.00 & 31.30 & \textbf{40.30} & \textbf{36.00} & 33.70 & 32.30 & 28.70 & 34.95 \\
Meta-Llama-3.1-70B-Instruct & 36.70 & 35.70 & 33.70 & 36.30 & 34.70 & 32.70 & 31.00 & 25.30 & 33.26 \\
\textbf{glm-zero-preview}$^\dagger$ & 34.30 & 28.70 & 27.30 & 32.70 & 26.00 & 33.70 & 27.30 & 25.70 & 29.46 \\
Qwen2.5-32B-Instruct & 34.70 & 31.30 & 30.00 & 29.00 & 25.30 & 34.00 & 24.30 & 23.30 & 28.99 \\
Qwen2.5-14B-Instruct & 30.00 & 28.70 & 26.00 & 30.00 & 24.70 & 31.70 & 24.00 & 25.00 & 27.51 \\
Meta-Llama-3.3-70B-Instruct & 31.30 & 27.70 & 26.30 & 27.00 & 26.00 & 26.70 & 23.70 & 25.30 & 26.75 \\
\textbf{Qwen-QwQ-32B-preview}$^\dagger$ & 26.70 & 24.70 & 26.30 & 30.00 & 22.70 & 26.30 & 24.70 & 27.70 & 26.14 \\
Meta-Llama-3.2-3B-Instruct & 22.70 & 20.00 & 22.00 & 25.00 & 21.00 & 25.70 & 23.70 & 20.70 & 22.60 \\
Meta-Llama-3.1-8B-Instruct & 22.00 & 22.00 & 22.00 & 22.30 & 24.00 & 24.00 & 24.00 & 20.30 & 22.57 \\
\textbf{Alibaba-Marco-o1}$^\dagger$ & 25.30 & 18.70 & 21.30 & 25.00 & 15.70 & 25.00 & 20.70 & 16.70 & 21.05 \\
Qwen2.5-7B-Instruct & 21.70 & 19.70 & 20.30 & 24.00 & 17.70 & 22.70 & 18.30 & 13.30 & 19.71 \\
Qwen2.5-3B-Instruct & 18.70 & 19.00 & 19.70 & 20.70 & 15.30 & 24.30 & 15.00 & 15.00 & 18.46 \\
Meta-Llama-3.2-1B-Instruct & 8.00 & 5.00 & 6.70 & 9.30 & 6.30 & 10.30 & 10.70 & 9.70 & 8.25 \\
\bottomrule
\end{tabular}
\end{table*}